\documentclass[twocolumn]{jpsj2}

\title{The ALPS project: open source software for strongly correlated systems}

\author{F. Alet$^{1,2,3}$, P. Dayal$^{1}$, A. Grzesik$^{4}$, A. Honecker$^{5}$, M. K\"orner$^{1}$,  A. L\"auchli$^{6}$, S.R. Manmana$^{7,8}$, I.P.~McCulloch$^{9}$, F.~Michel$^{10}$, R.M. Noack$^{8}$,  G. Schmid$^{1}$, U. Schollw\"ock$^{9}$, F. St\"ockli$^{1}$, S. Todo$^{11}$, S. Trebst$^{1,2}$, M. Troyer$^{1,2}$, P.~Werner$^{1}$, and S.~Wessel$^{1,7}$  (ALPS collaboration)}
\inst{
$^{1}$ Theoretische Physik, ETH Z\"urich, CH-8093 Z\"urich, Switzerland \\
$^{2}$ Computational Laboratory, ETH Z\"urich, CH-8092 Z\"urich, Switzerland \\
$^{3}$ Service de Physique Th{\'e}orique, CEA Saclay, F-91191 Gif sur Yvette, France \\
$^{4}$ Physikalisches Institut der Universit\"at Bonn, D-53115 Bonn, Germany\\
$^{5}$ Institut f\"ur Theoretische Physik, Technische Universit\"at
            Braunschweig, D-38106 Braunschweig, Germany\\
$^{6}$ IRRMA, EPF Lausanne, PPH-Ecublens, CH-1015 Lausanne, Switzerland \\
$^{7}$ Institut f\"ur Theoretische Physik III, Universit\"at Stuttgart, Pfaffenwaldring 57, D-70550 Stuttgart, Germany \\
$^{8}$ AG Vielteilchennumerik, Fachbereich Physik, Philipps-Universit\"at Marburg, D-35032 Marburg, Germany \\
$^{9}$ Institute f\"ur Theoretische Physik C, RWTH Aachen, D-52056 Aachen, Germany \\
$^{10}$ Institut f\"ur Theoretische Physik, Technische Universit\"at Graz, Petersgasse 16, A-8010 Graz, Austria \\
$^{11}$ Department of Applied Physics, University of Tokyo, 113-8656 Tokyo, Japan
}
\date{\today}

\abst{

We present the ALPS (Algorithms and Libraries for Physics Simulations)
project, an international open source software project to develop
libraries and application programs for the simulation of strongly
correlated quantum lattice models such as quantum magnets, lattice
bosons, and strongly correlated fermion systems. Development is
centered on common XML and binary data formats, on libraries to
simplify and speed up code development, and on full-featured
simulation programs. The programs enable non-experts to start carrying
out numerical simulations by providing basic implementations of the
important algorithms for quantum lattice models: classical and quantum
Monte Carlo (QMC) using non-local updates, extended ensemble
simulations, exact and full diagonalization (ED), as well as the
density matrix renormalization group (DMRG). The software is available
from our web server at {\tt http://alps.comp-phys.org/}.}

\kword{quantum lattice model, open source software, C++, Monte Carlo, quantum Monte Carlo, density matrix renormalization group, exact diagonalization}

\begin{document}
\maketitle

\section{Introduction} 

Quantum fluctuations and competing interactions in quantum many body
systems lead to unusual and exciting properties of strongly correlated
materials such as quantum magnetism,\cite{QMbook} high temperature
superconductivity,\cite{BednorzMueller} heavy fermion
behavior,\cite{HeavyFermion} and topological quantum order.\cite{Wen}
Although our understanding of these intriguing systems has grown over
the past two decades, many questions remain unanswered. Strong
interactions, competition between various ordered states, and strong
thermal and quantum fluctuations impose limits on the applicability of
mean-field theories and perturbative approaches. While renormalization
group (RG) approaches can predict scaling relations in these strongly
correlated regimes, the validity for realistic microscopic models of
the effective actions used in the RG approaches can only be verified
by numerical methods.

Direct numerical simulations of microscopic models for strongly
correlated systems using unbiased and accurate algorithms are
therefore of growing importance for the understanding of the unusual
properties of these systems. Given the challenges faced by classical
many-body simulations, such as critical slowing down and tunneling
problems in classical Monte Carlo simulations\cite{LandauBinder}, it
is no surprise that simulations of quantum many-body systems present
even larger challenges. While the problems of critical slowing down and of
tunneling through energy barriers at first order phase transitions
have been solved for bosonic and non-frustrated magnetic systems by a
generalization of the classical cluster Monte
Carlo\cite{cluster} and extended ensemble (broad histogram)
algorithms to quantum systems,\cite{QWL} the simulation of fermionic
systems still presents major difficulties. 
Indeed, it has recently been shown that the negative sign
problem of fermionic quantum Monte Carlo simulations (QMC) is
NP-hard,\cite{nphard} and that a general algorithm with polynomial
scaling most likely does not exist. The most promising
approach therefore appears to 
be to improve and use a variety of complementary algorithms, such as
exact diagonalization\cite{ED}, series expansion\cite{series} or the
density matrix renormalization group (DMRG) 
method\cite{DMRG,DMRGreview,dmrgbook}.

The last decade has seen tremendous progress in the development of
algorithms.  Speedups of many orders of magnitude have been 
achieved.\cite{cluster,QWL,DMRGreview,dmrgbook} These
advances come at the cost of substantially increased algorithmic
complexity and challenge the current model of program development in
this research field. In contrast to other research areas, in which
large ``community codes'' are being used, the field of strongly
correlated systems has so far been based mostly on single codes developed by
individual researchers for particular projects. While the simple
algorithms used a decade ago could be easily programmed by a beginning
graduate student in a matter of weeks, it now takes substantially
longer to master and implement the new algorithms.  Thus, their use
has increasingly become restricted to a small number of experts.

\section{The ALPS project}
The ALPS (Algorithms and Libraries for Physics Simulations) project aims to
overcome the problems posed by the growing complexity of algorithms
and the specialization of researchers onto single algorithms through
an open-source software development initiative. Its goals are to
provide:
\begin{itemize}
\item {\bf standardized file formats} to simplify exchange,
distribution and archiving of simulation results and to achieve
interoperability between codes.
\item {\bf generic and optimized libraries} for common aspects of
simulations of quantum and classical lattice models, to simplify code
development.
\item a set of {\bf applications} covering the major algorithms.
\item{\bf license} conditions that encourage researchers to contribute
to the ALPS project by gaining scientific credit for use of their
work.
\item {\bf outreach} through a web page,\cite{alps} mailing lists and
workshops to distribute the results and to educate researchers both
about the algorithms and the use of the applications.
\item {\bf improved reproducibility} of numerical results by
publishing source codes used to obtain published results.
\item an {\bf archive} for simulation results.
\end{itemize}

The ready-to-use applications are useful both for 
{\it theoreticians} who want to test theoretical ideas about quantum
lattice models and to explore their properties, as well as for 
{\it experimentalists} trying to fit experimental data to theoretical
models to obtain information about the microscopic properties of
materials.

In contrast to other fields, in which open-source development efforts
began decades ago, starting now allows us to build our design directly
on recent developments in computer science, in particular: 
\begin{itemize}
\item XML\cite{xml} (eXtensible Markup Language) and HDF5\cite{hdf5}
(Hierarchical Data Format version 5), portable data formats
supported by a large number of standard tools
\item generic and object oriented programming in C++ to achieve
flexible but still optimal codes
\item OpenMP\cite{openmp} and MPI\cite{mpi} for parallelization on
shared memory machines and Beowulf clusters
\end{itemize} 

 In the following, we present additional details about these various
 aspects of the ALPS project. 
\section{File formats}

The most basic part of the ALPS project is the definition of
common standardized file formats suitable for a wide range of
applications. Standardized file formats enable the exchange of data
between applications, allow the development of common evaluation
tools, simplify the application of more than one algorithm to a given
model, and are a prerequisite for the storage of simulation data in a
common archive.

\begin{figure}[tb]
\begin{small}
\begin{verbatim}
<PARAMETERS>
  <PARAMETER name="LATTICE"> square </PARAMETER>
  <PARAMETER name="MODEL">     spin </PARAMETER>
  <PARAMETER name="L">           10 </PARAMETER>
  <PARAMETER name="T">          0.5 </PARAMETER>
</PARAMETERS>
\end{verbatim}
\end{small}
\caption{Excerpt from an XML file for simulation input parameters.}
\label{fig:parm}
\end{figure}

\begin{figure}[tb]
\begin{small}
\begin{verbatim}
<SCALAR_AVERAGE name="Susceptibility">
  <MEAN>                     421.3 </MEAN>
  <ERROR converged="yes">     1.54 </ERROR>
  <VARIANCE>              1.06e+05 </VARIANCE>
</SCALAR_AVERAGE>
\end{verbatim}
\end{small}
\caption{Excerpt from an XML file for simulation results of the uniform susceptibility in a Monte Carlo simulation.}
\label{fig:result}
\end{figure}

We have designed a number of XML schemas\cite{xmlschema} to describe
the input of simulation parameters, the output of results, and the
specification of lattices and quantum lattice models.  The ISO
standard XML was chosen for the specification because it has become
the main text-based data format on the internet and because it is
supported by a large and growing number of tools.  Unlike simple
formats, in which the location of an argument in the file specifies its
meaning (e.g. the first line specifies the system size, the second line
temperature), XML specifies the meaning of data through
meta-information provided by markup with tags, as shown in figure
\ref{fig:parm} for input parameters and figure \ref{fig:result} for
simulation results. The meaning of these files is easy to decipher
even years after completion of the simulation, unlike many other
formats. The resemblance to HTML is not accidental; in fact, HTML
(XHTML) is also an XML format.

\begin{figure}
\begin{small}
\begin{verbatim}
<LATTICEGRAPH name = "square">
  <FINITELATTICE>
    <LATTICE dimension="2"/>  
    <EXTENT dimension="1" size="L"/>
    <EXTENT dimension="2" size="L"/>
    <BOUNDARY type="periodic"/>  
  </FINITELATTICE>
  <UNITCELL>
    <VERTEX/>
    <EDGE>
      <SOURCE vertex="1" offset="0 0"/>
      <TARGET vertex="1" offset="0 1"/>
    </EDGE>
    <EDGE>
      <SOURCE vertex="1" offset="0 0"/>
      <TARGET vertex="1" offset="1 0"/>
    </EDGE>
  </UNITCELL> 
</LATTICEGRAPH>
\end{verbatim}
\end{small}
\caption{The definition of a square lattice with one site (vertex) per
unit cell and bonds (edges) only to nearest neighbors. First the
dimension, extent, and boundary conditions of the Bravais lattice are
described in the {\tt <FINITELATTICE>} element, then the unit cell
including the bonds in the lattice is defined.}
\label{fig:lattice}
\end{figure}

\begin{figure}
\begin{small}
\begin{verbatim}
<BASIS name="spin">
  <SITEBASIS>
    <QUANTUMNUMBER name="S" min="1/2" max="1/2"/>
    <QUANTUMNUMBER name="Sz" min="-S" max="S"/>
    <OPERATOR name="Splus" 
        matrixelement="sqrt(S*(S+1)-Sz*(Sz+1))">   
      <CHANGE quantumnumber="Sz" change="1"/>
    </OPERATOR>
    <OPERATOR name="Sminus" 
        matrixelement="sqrt(S*(S+1)-Sz*(Sz-1))">  
      <CHANGE quantumnumber="Sz" change="-1"/>
    </OPERATOR>
    <OPERATOR name="Sz" matrixelement="Sz"/>  
  </SITEBASIS>
</BASIS>

<HAMILTONIAN name="spin">
  <BASIS ref="spin"/>
  <SITETERM> -h*Sz </SITETERM>   
  <BONDTERM source="i" target="j">
    Jxy/2*(Splus(i)*Sminus(j)+Sminus(i)*Splus(j))
    + Jz*Sz(i)*Sz(j)
  </BONDTERM>
</HAMILTONIAN>
\end{verbatim}
\end{small}
\caption{The definition of a spin-1/2 $XXZ$ spin Hamiltonian with one
  type of exchange coupling only: 
  $H=-h\sum_iS_i^z+\sum_{\langle
    i,j\rangle} \left((J_{\rm
    xy}/2)(S_i^+S_j^-+S_i^-S_j^+)+J_zS_i^zS_j^z\right)$.  After
  describing the local basis for each site and operators acting on it,
  the Hamiltonian is defined.}
\label{fig:model}
\end{figure}

In addition to XML schemas for parameter input and for output of results,
we have developed schemas for the description of lattices and
models. Examples are shown in figures \ref{fig:lattice} and
\ref{fig:model}.  A detailed specification of the formats is given on
our XML web page.\cite{xmlschema}

Any of the ALPS applications can be run by providing an input file
(such as the one which is excerpted in figure \ref{fig:parm}),
together with lattice and model definitions (figures \ref{fig:lattice}
and \ref{fig:model}) to that application (provided the application
supports that type of model).  The application then returns an output
file containing data such as that shown in figure \ref{fig:result}.
Standardized formats that extend across all applications reduce the
learning curve for using the applications and allow common tools to be
used to analyze the data.

In order to avoid working directly with the long (and sometimes ugly)
output files in XML format, the XML files can be easily transformed to
other formats using XSLT transformations\cite{xslt} and can be viewed,
for example, directly as HTML in a web browser, printed as plain text,
converted to \LaTeX, or displayed as a plot in one of a number of
common plotting programs.\cite{plot}

In addition to these text-based XML formats, the ALPS libraries
support portable binary formats for large data sets such as time
series of Monte Carlo simulations. These data sets are currently
stored in XDR format, the standard Unix format for remote procedure
calls and serialization, but plans are afoot to switch to the
better-supported HDF5 format\cite{hdf5} in the near future.

\section{Libraries}
The ALPS libraries are the foundation of all the ALPS applications,
providing functionality common to all of them:
\begin{itemize}
\item an XML parser and output stream to read and write the ALPS XML
files.
\item an expression library to manipulate and evaluate symbolic
expressions.
\item a scheduler for the automatic parallelization of Monte Carlo
simulations and other ``embarrassingly'' parallel applications,
implementing load balancing and checkpointing.\cite{palm}
\item a lattice structure library for the creation of arbitrary graphs
and Bravais lattices from XML input.
\item a model library for the construction of basis sets, operators,
and Hamiltonians from XML input.
\item the ``alea'' library for the statistical analysis and evaluation
of Monte Carlo data including a binning analysis of errors and
jackknife analysis of cross-correlations.\cite{palm}
\item a serialization library ``osiris'' for the serialization of C++
data structures, used for writing program checkpoints and portable
binary result files.\cite{palm}
\end{itemize}
These libraries make full use of object-oriented and generic
programming techniques,\cite{CE} which allows them to be very flexible without
losing any performance compared to FORTRAN programs. To give just one
example, the lattice library, which is implemented generically using
C++ template features and is based on the Boost Graph
Library\cite{BGL,boost} (BGL), does not restrict the user to a
specific data structure as in C or Fortran libraries. Instead, the
application programmer can choose the data structure best suited for
the application.  As long as this data structure provides the BGL
graph interface, the ALPS lattice library can construct the lattice
from the XML description.

\begin{figure}
\begin{center}
\includegraphics[width=8cm]{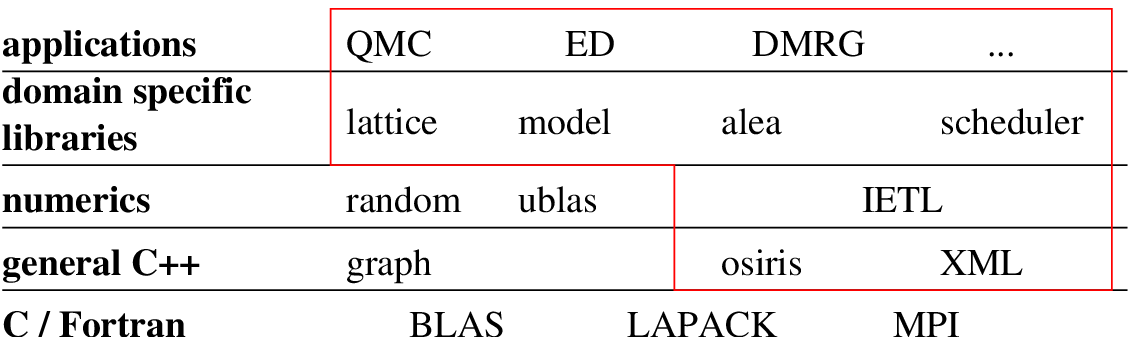}
\end{center}
\caption{Hierarchical structure of the components of the ALPS
project. The red frame encloses components developed as part of the
ALPS project. Libraries outside the frame are publicly available
high-performance libraries that can be obtained from the
Boost\cite{boost} web page or the netlib repository.\cite{netlib}}
\label{fig:libs}
\end{figure}

Figure \ref{fig:libs} shows a diagrammatic overview of the hierarchy
of components in the ALPS project: the ALPS applications (discussed in
the next section) are based on domain-specific ALPS libraries, such as
the lattice, model, alea, or the scheduler library. These libraries
are based on numerical and general C++ libraries, developed either in
ALPS or in the C++ Boost project \cite{boost}. In addition to the BGL
mentioned above, the external libraries include the Boost random
number library, which has been accepted into the next release of the
C++ standard as a standard library, and the Boost uBLAS library for
matrix and vector operations. A numerical library developed by us and
distributed with the ALPS software, the Iterative Eigensolver Template
Library\cite{ietl} (IETL), provides a package of iterative
eigensolvers for sparse matrices. Finally, the lowest level of the
hierarchy is a set of common C and FORTRAN libraries, including the
high performance BLAS and LAPACK libraries for linear algebra and the
MPI and PVM libraries for communication on distributed systems.

\section{Applications}
\label{sec:applications}
The following application programs, based on the above libraries, are
implemented in release 1.2:
\begin{itemize}
\item a Monte Carlo program for classical magnets employing cluster
algorithms.\cite{SW}
\item the QMC program ``looper'' using the loop cluster
algorithm\cite{cluster,looper} in a stochastic series
expansion\cite{sse} (SSE) and path-integral representation for quantum
magnets with spin reversal symmetry.
\item an optimized directed loop QMC program\cite{directed} in an SSE
representation for bosonic and magnetic quantum lattice models.
\item a worm algorithm\cite{worm} QMC program best suited for bosonic
quantum lattice models and easy-axis quantum magnets.
\item an extended ensemble QMC program using a generalization of the
Wang-Landau\cite{WangLandau} algorithm to quantum systems,\cite{QWL}
to obtain thermodynamic quantities over large temperature ranges.
\item an exact diagonalization program which can calculate the ground state
and low lying excited states of quantum lattice models using the
Lanczos\cite{lanczos} algorithm.
\item a full diagonalization program which can calculate the complete
eigenspectrum of quantum lattice models and from it all thermodynamic
properties.
\item a DMRG\cite{DMRG} program for {\em noninteracting}
particles. This is an adaptation of the program 
in Refs.\ \cite{dmrgbook,noninteracting} to the ALPS libraries. 
In addition to demonstrating the fundamental ideas of the DMRG, it solves
one-dimensional single-particle quantum problems in arbitrary
potentials.
\end{itemize}

Planned for the future are:
\begin{itemize}
\item support for lattice symmetries in the exact and full diagonalization codes in release 1.3.
\item a DMRG\cite{DMRG,DMRGreview,dmrgbook} code for static properties in release 1.3.
\item a series expansion\cite{series} code for both perturbation
series and high temperature series.
\item a determinental QMC\cite{determinental} code for two- and
three-dimensional fermionic models.
\item a toolset for mean-field calculations on quantum lattice models.
\end{itemize}

These applications comprise the most important unbiased
algorithms for quantum many body systems. The implementations all
share the same file formats, simplifying their use, reducing the
learning curve, and enabling the easy investigation of a model with
more than one method. Tutorials on the use of the applications are
included with the sources that can be found on the ALPS web
page.\cite{alps}

These codes and libraries have already been used in a number of
publications, of which we will mention only a small representative
sample illustrating the breadth of possible applications.  The exact
diagonalization library was used in the investigation of quantum bits
with topological protection.\cite{qubit} Anisotropic quasi-one
dimensional and quasi-two dimensional quantum magnets were
investigated recently using the ``looper'' QMC program.\cite{quasi} A
precursor of the looper program was used for extensive fits of QMC
simulations to experimental measurements\cite{Johnston,experiments}
and to determine phase diagrams critical exponents of quantum phase
transitions\cite{QPT} and universal critical temperatures of quantum
critical magnets in a field.\cite{KT} A precursor of the directed loop
program based on the ALPS libraries has been used to determine ground
state and finite-temperature phase diagrams of hard-core
bosons.\cite{hcbosons} The optimized directed loop and worm algorithm
QMC programs were applied to determine properties of ultra-cold
bosonic atoms in optical lattices\cite{boson} and to calculate
magnetization curves of spin systems.\cite{magn} The ALPS libraries
were also employed recently in simulations of dissipative quantum
phase transitions\cite{dissipation}, in a spin-wave analysis of
quantum magnetic quasicrystals,\cite{quasicrystal} and in a
multi-scale approach to the simulation of nanomagnets.\cite{multi}

\section{License and Motivations to Contribute}
The ALPS library\cite{liblicense} and
application\cite{applicationlicense} licenses were designed by
I.P. McCulloch after extensive discussions inside the ALPS
collaboration. They are based on the GNU general public
license.\cite{gnulicense} The use of the libraries and codes is free
but carries a citation requirement to publications describing the
libraries and codes.

Stricter licensing terms, such as the requirement to submit any
modifications and improvements of the codes to the ALPS collaboration,
or a requirement to publish any codes built on the ALPS libraries or
applications as a supplement to the paper presenting the results, were
discussed but have not been included in the license at the present
time. We feel, however, that the idea of publishing source code of
numerical simulations together with the scientific results produced
could be very beneficial by enabling reproducibility of numerical
simulations.  We encourage a wider discussion on this question in the
computational physics community.

Besides ``payment'' in the scientific currency of citations, and peer
pressure to contribute after profiting from the ALPS project,
additional motivations (especially for junior researchers) to
contribute to the ALPS project are networking opportunities and many
chances to start new research collaborations based on codes
contributed to the ALPS project.

\section{Outreach and Development}
The ALPS web page\cite{alps} is the central information
repository of the ALPS collaboration.  It makes source codes of the
ALPS libraries and applications available, provides access to the
documentation, and distributes information about the mailing
lists, new releases, and workshops.

We intend to provide extensive tutorials with hands-on sessions using
the ALPS applications at workshops on numerical simulations of quantum
lattice models at least once per year. Such tutorials have thus far
been held on three occasions: a workshop on Wang-Landau sampling for
classical and quantum systems at Oak Ridge National Labs in August
2003, a graduate lecture series at the University of Hannover, Germany
in November 2003, and a workshop on simulation tools in
Lugano, Switzerland in September 2004.\cite{lugano} More tutorials are
planned for 2005 and 2006.

The ALPS project has been further publicized by talks, such as an invited
talk by S. Trebst at the 2004 March Meeting of the American Physical
Society,\cite{trebstaps} a presentation at the DMRG workshop at the
Lorentz center in Leiden,\cite{noackleiden} and poster presentations at
various conferences.

Development of the ALPS libraries and applications is coordinated
through both mailing lists and semi-annual developer workshops. These
ALPS developer workshops (all held in alpine regions so far) have been
instrumental not only in planning development efforts for the subsequent
half year, but also in providing important development deadlines -- and
thus counter the susceptibility of all software projects to
interminable delays.

\section{Future plans}

The ALPS project is continuously evolving; many improvements and
additions are planned for forthcoming releases.  Planned near-term
improvements, in addition to those mentioned in
section \ref{sec:applications} will be the capability of specifying
arbitrary measurements of local operators and correlation functions in
input files.

A major future development effort will be to establish a database and
archive for simulation results based on our proposed XML
formats. Whereas today simulation results are usually published only
in the form of figures or fit results after much data processing, the
archive will allow the easy retrieval of detailed results in a unified
format. Current research on data and storage grids as well as the
rapid progress in the development of XML-based databases and tools
will surely prove helpful for this project.

In a further stage, we envision expanding the ALPS project in the long
term to include dynamical mean field theory (DMFT) codes\cite{dmft} as
well as interfacing the ALPS applications to band structure codes,
thus making it possible to obtain a microscopic lattice model from
ab-initio calculations. These efforts will be coordinated with the
$\psi$-Mag project at Oak Ridge National Laboratory, which aims at
unifying the interface of band structure codes and at combining them with
classical spin dynamics simulations for nanomagnetics.\cite{ornl}

In addition to these coordinated development efforts, we offer a
repository for other open source simulation codes on our computational
physics web page {\tt http://www.comp-phys.org/}.

\section{Conclusions}

The ALPS project is an open source effort at providing libraries and
applications for the simulation of classical and quantum lattice
models. Common XML file formats enable the sharing and archiving of
simulation data. Generic and optimized libraries simplify the
development of simulation codes by implementing common tasks such as
I/O and the construction of lattice and model data structures. The
ALPS applications make modern high-performance numerical algorithms
available to a wider range of researchers. They enable theoreticians
to investigate properties of interesting strongly correlated models
conveniently, and allow experimentalists to perform direct comparisons
and fits of experimental measurements to numerical simulations.

Researchers interested in announcements of new releases, information
about workshops, or in contributing to the ALPS project are encouraged
to sign up to the mailing lists on our web page.\cite{alps}

\section*{History and Acknowledgment}

The ALPS project was formally initiated at a workshop in Guarda,
Switzerland in January 2003,\cite{guarda} by merging the PALM++
library\cite{palm} for distributed Monte Carlo simulations and
projects based on it, the open-source Looper\cite{looper} code, and
the IETL library for sparse eigensolvers,\cite{ietl} with development
efforts on DMRG and other algorithms.

The initial idea for the project was seeded by D.C. Johnston (Ames
Laboratory), who in the course of fruitful collaborations on fitting
experimental measurements on quasi-one dimensional systems to QMC
simulations,\cite{Johnston} demonstrated the usefulness of
computational algorithms for experimental physicists. Additional
motivation came from our desire to ease the exchange of codes in
research collaborations, and by the $\psi$-Mag project at Oak Ridge
National Laboratories.\cite{ornl}

We acknowledge support by the Swiss National Science Foundation, the
Kavli Institute for Theoretical Physics at the University of
California at Santa Barbara, and the Aspen Center for Physics. We are
grateful to many experimental physicists, especially to D.C. Johnston,
C.P. Landee, and V. Tangoulis for making us aware of the interest of
experimentalists in numerical simulations of quantum systems and for
their encouragement to start this project, and to H.G. Evertz for
advice on the implementation of error evaluation in Monte Carlo
simulations.


\begin{thebibliography}{99} 
\bibitem{QMbook} U. Schollw\"ock {\it et al.} (Eds.): {\it Quantum magnetism}, Lecture Notes in Physics {\bf 645} (Springer Verlag, 2004).
\bibitem{BednorzMueller} J. G. Bednorz and K. A. M\"uller: Z. Phys. B {\bf 64} (1986) 189.
\bibitem{HeavyFermion} for a review see P. Thalmeier, G.  Zwicknagl, O. Stockert, G. Sparn, and F. Steglich:
  Report cond-mat/0409363.
\bibitem{Wen} for a review see X.G. Wen: {\it Quantum Field Theory Of Many-body Systems: From The Origin Of Sound To An Origin Of Light And Electrons } (Oxford University Press, 2004)
\bibitem{LandauBinder} D. P. Landau and K. Binder:
{\it A Guide to Monte Carlo Simulations in Statistical Physics} 
(Cambridge University Press, Cambridge, October 2000).
\bibitem{cluster} For a review see M. Troyer {\it et al.}: AIP Conf. Proc. {\bf 690} (2003) 156; H.G. Evertz, Adv. in Physics {\bf 52} (2003) 1.
\bibitem{QWL} M. Troyer, S. Wessel, and F. Alet: Phys. Rev. Lett. {\bf 90} (2003) 120201.
\bibitem{nphard} M. Troyer and U.-J. Wiese: Report cond-mat/0408370.
\bibitem{ED} for a review see N. Laflorencie and D. Poilblanc: Lect. Notes Phys. {\bf 645} (2004) 227.
\bibitem{series} M. P. Gelfand and R. R. P. Singh: Advances in Physics {\bf 49} (2000) 93.
\bibitem{DMRG} S.R. White: Phys. Rev. Lett. {\bf 69} (1992) 2863; Phys. Rev.
B {\bf 48} (1993) 10345.
\bibitem{DMRGreview} for a review see: U. Schollw\"ock: Report cond-mat/0409292, to be published in Rev. Mod. Phys. (2005).
\bibitem{dmrgbook} I. Peschel  {\it et al.} (Ed.): {\it Density Matrix Renormalization - A New Numerical Method in Physics}, (Springer Verlag, Berlin, 1999).
\bibitem{alps} {\tt http://alps.comp-phys.org/}
\bibitem{xml} {\tt http://www.w3.org/XML/}
\bibitem{hdf5} {\tt http://hdf.ncsa.uiuc.edu/HDF5/}
\bibitem{openmp} {\tt http://www.openmp.org/}
\bibitem{mpi} {\tt http://www.mpi-forum.org/}
\bibitem{xmlschema} See {\tt http://xml.comp-phys.org/} for the XML schemas and their documentation as well as mailing lists.
\bibitem{xslt} {\tt http://www.w3.org/TR/xslt/}
\bibitem{plot} Currently we directly support Grace and Gnuplot formats, but other formats can easily be added.
\bibitem{palm} M. Troyer, B. Ammon, and E. Heeb: Lecture Notes in Computer Science {\bf 1505} (1998) 191.
\bibitem{CE} K. Czarnecki and U.W. Eisenecker: {\it Generative Programming}, (Addison Wesley, 2000).
\bibitem{BGL} J.G. Siek, L.-Q. Lee, A. Lumsdaine: {\it The Boost Graph Library User Guide and Reference Manual} (Addison-Wesley, 2001).
\bibitem{boost} {\tt http://www.boost.org/}
\bibitem{netlib} {\tt http://www.netlib.org/}
\bibitem{ietl} {\tt http://www.comp-phys.org/software/ietl/}
\bibitem{SW} R.H. Swendsen and J-S. Wang: Phys. Rev. Lett {\bf 58} (1987) 86; U. Wolff, Phys. Rev. Lett {\bf 62} (1989)  361.
\bibitem{looper} S. Todo and K. Kato, Rev. Lett. {\bf 87} (2001) 047203. {\tt http://wistaria.comp-phys.org/alps-looper/}
\bibitem{sse} A.W. Sandvik and J. KurkijŠrvi: Phys. Rev. B {\bf 43} (1991) 5950; A.W. Sandvik: Phys. Rev. B {\bf 59} (1999) R14157.
\bibitem{directed} F. Alet, S. Wessel, and M. Troyer: Report cond-mat/0308495.
\bibitem{worm} N.V. Prokof'ev, B.V. Svistunov, and I.S. Tupitsyn: Sov. Phys. JETP {\bf 87} (1998) 310.
\bibitem{WangLandau} F. Wang and D.P. Landau: Phys. Rev. Lett. {\bf 86} (2001) 2050; Phys. Rev. E {\bf 64} (2001) 056101. 
\bibitem{lanczos} C. Lanczos: J. Res. Natl. Bur. Stand. {\bf 45} (1950) 225. 
\bibitem{noninteracting} M.A. Martin-Delgado, G. Sierra, and R. M. Noack: J. Phys. A {\bf 32} (1999) 6079;
 and chapter 2 of Ref. \citen{dmrgbook}.
\bibitem{determinental} F. F. Assaad in: J. Grotendorst, D. Marx, and A. Muramatsu (Eds): {\it Lecture notes of the Winter School on Quantum Simulations of Complex
Many-Body Systems: From Theory to Algorithms}, Publication Series
of the John von Neumann Institute for Computing (NIC). NIC series Vol. 10.
\bibitem{qubit} L. Ioffe {\it et al}: Nature {\bf 415} (2002) 507; M.V. Feigel'man {\it et al.}: Phys. Rev. Lett. {\bf 92} (2004) 098301.
\bibitem{quasi} C. Yasuda {\it et al.}: Report cond-mat/0312392.
\bibitem{Johnston} D.C. Johnston {\it et al.}: Phys. Rev. B {\bf 61} (2000) 9558; D.C. Johnston {\it et al.}: Report cond-mat/0001147.
\bibitem{experiments} R. Melzi {\it et al.}: Phys. Rev. Lett. {\bf 85} (2000) 1318; B. Pedrini {\it et al.}: Report cond-mat/0402482.
\bibitem{QPT} M. Troyer, H. Kontani, and K. Ueda: Phys. Rev. Lett. {\bf 76} (1996) 3822;
M. Troyer, M. Imada, and K. Ueda: J. Phys. Soc. Jpn. {\bf 66} (1997) 2957.
\bibitem{KT} M. Troyer and S. Sachdev: Phys. Rev. Lett. {\bf 81} (1998) 5418.
\bibitem{hcbosons} F. Hebert {\it et al.}: Phys. Rev. B {\bf 65} (2002) 014513; A. Dorneich {\it et al.}: Phys. Rev. Lett. {\bf 88} (2002) 057003; K. Bernardet {\it et al.}: Phys. Rev. B {\bf 65} (2002) 104519; 
G. Schmid {\it et al.}: Phys. Rev. Lett. {\bf 88}, 167208 (2002); K. Bernardet {\it et al.}: Phys. Rev. B {\bf 66} (2002) 054520; G. Schmid and M. Troyer: Phys. Rev. Lett. {\bf 93} (2004) 067003.
\bibitem{boson} S. Wessel {\it et al.}: Advances in Solid State Physics {\bf 44} (2004) 265; Report cond-mat/0404552, Phys. Rev. A (2004), in press; see also the contribution in this volume.
\bibitem{magn}  J. Richter, J. Schulenburg, and A. Honecker:
  Lect. Notes Phys. {\bf 645} (2004) 85. 
\bibitem{dissipation} S. Sachdev, P. Werner, and M. Troyer: Phys. Rev. Lett. {\bf 92} (2004) 237003;
P. Werner, K. V\"olker, M. Troyer, and S. Chakravarty: Report cond-mat/040224;
P. Werner and M. Troyer: Report cond-mat/0409664;
P. Werner, M. Troyer, and S. Sachdev, contribution in this volume.
\bibitem{quasicrystal} S. Wessel and I. Milat: Report cond-mat/0410180.
\bibitem{multi} S. Wessel, S. Trebst, and M. Troyer: {\it A multiscale approach to the simulation of quantum effects in nano-scale magnetic systems}, SIAM Journal on Multiscale Modeling and Simulation (2004), in print.
\bibitem{liblicense}  {\tt http://alps.comp-phys.org/software/ALPS/LICENSE.txt}
\bibitem{applicationlicense} {\tt http://alps.comp-phys.org/software/applications/ LICENSE.txt}
\bibitem{gnulicense}  {\tt http://www.gnu.org/software/}
\bibitem{lugano} {\tt http://www.comp-phys.org/lugano04/}
\bibitem{trebstaps} S. Trebst: Bull. Am. Phys. Soc. {\bf 49} (2004) J5.002 ; {\tt http://www.aps.org/meet/MAR04/baps/abs/S3250002.html}
\bibitem{noackleiden} {\tt http://dmrg.info/workshop/}
\bibitem{dmft} A. Georges {\it et al.}: Rev. Mod. Phys {\bf 68} (1996) 13; T. Pruschke, M. Jarrell, and J.K. Freericks: Adv. Phys. {\bf 42} (1995) 187.
\bibitem{ornl} {\tt http://mri-fre.ornl.gov/psimag}
\bibitem{guarda} {\tt http://alps.comp-phys.org/resources/workshops/
d0103/index.html}
\end{thebibliography}
\end{document}